\documentclass[11pt]{article}
\usepackage{epsf}
\usepackage{graphicx}        
\input epsf
\def\ll{\label}

\def\r1{(\ref{$1})}

\def\ba{\begin{array}{c}}

\def\ea{\end{array}}

\def\De{\Delta}

\def\l{\left}
\def\l({\left(}
\def\r){\right)}
\def\r{\right}

\def\la{\lambda}

 \def\be{\begin{equation}}
\def\bc{\begin{center}}
\def\ec{\end{center}}
\def\bit{\begin{itemize}}
\def\eit{\end{itemize}}
\def\ee{\end{equation}}
\def\ed{\end{document}}
\def\bea{\begin{eqnarray}}
\def\eea{\end{eqnarray}}
\def\efr{\end{flushright}}

\begin{document}

\title{{ Construction and  
exact solution of a  nonlinear  quantum field model   
in quasi-higher dimension
 }}
\author{Anjan Kundu\\
Theory
Division, Saha Institute of Nuclear Physics, 1/AF, Bidhannagar,\\ Kolkata
700064, India
\\{\it email}: anjan.kundu@saha.ac.in\ \ {\it phone}:
+91-9433021522 \ \ {\it FAX}:   +91-3323374637}
\vskip 1cm 




 \maketitle

\begin{abstract}
Nonperturbative exact solutions are allowed for 
  quantum  integrable models  in one space-dimension.  
 Going beyond this class we propose an alternative Lax matrix approach,
  exploiting the hidden multi-space-time concept in integrable systems and
 construct a  novel   nonlinear Schr\"odinger quantum  field   model in quasi-two
dimensions.  An intriguing field commutator is discovered, confirming the
integrability of the model and yielding its
   exact Bethe ansatz solution with   rich scattering and  bound-state
properties.  The universality of the  scheme is expected to cover diverse
models, opening up a new direction in the field.
 \end{abstract}
\noindent {PACS numbers}:
{ 02.30.lk,
03.65.Fd,
03.70.+k,
11.15.Tk

\vskip .3cm 

\noindent{MSC numbers}:
\ 82B23,
16T25,
81R12,
81R15
\vskip .7cm

\noindent {\bf keywords}:
 Higher order Lax operator;
quasi-two  dimensional nonlinear quantum field model; novel field
commutation algebra;  quantum  Yang-Baxter equation;  algebraic Bethe
ansatz; exact solution.
\maketitle

\section{Introduction and Motivation} A large number  of  quantum
models  in one space-dimension (1D)
   admits   exact nonperturbative solutions, in spite of their nonlinear interaction. 
This exclusive class of models, which also  includes    field models,  
constitute the family of quantum  integrable (QI) systems
\cite{fadrev,kulskly,korbook,baxter,MattisEncyl} with 
 extraordinary   properties, like association with a quantum  Lax 
and a quantum $R$ matrix, possessing rich  underlying  algebraic structures to
  satisfy the quantum Yang-Baxter equation \ \ (QYBE), existence of a  commuting set of
 conserved operators with an exact solution of their eigenvalue
problem (EVP), etc.  This exact  method of solution, known as the Bethe
ansatz (BA),  was  pioneered by Bethe way back in 1931 \cite{bethe31} and generalized
later to  algebraic BA \cite{fadrev,kulskly,korbook,baxter}.
 These  QI systems defined  in $1+1$-dimensions, include a wide variety of
  models, e.g.
 isotropic \cite{bethe31} and
anisotropic  \cite{xyz,xxz} quantum spin-$\frac 1 2 $ chains,
 $\delta $ and $\delta^{'} $-function Bose \cite{dbose,d1bose}
and anyon
 \cite{dany,d1any} gases,  nonlinear Schr\"odinger (NLS) field
\cite{fadrev,sklyaninNLS} and  lattice  \cite{korbook}
model, relativistic \cite{kunToda} and
 nonrelativistic \cite{toda} Toda chain,   t-J \cite{tj} and
Hubbard  \cite{hubb,hubbS} model,  \  Gaudin  model \cite{sklyGodin}, 
 derivative NLS
 \cite{kunDNLS}, sine-Gordon \cite{fadSG} and  Liuoville \cite{fadLiu}  model,
etc.  The algebraic structures  underlying these models are also rich and
diverse, which include canonical, bosonic, fermionic, anyonic and spin
algebras,
 quantum oscillator and quantum group algebras etc., having  inherent Hopf
algebra properties \cite{chari}.
  However, it is  important to note, that among this diversity there is a
deep unity, revealing that all known QI models, we are interested in, are
realizable from a single ancestor Lax matrix or its q-deformation
\cite{korbook,kunduPRL99}.  At the
same time, the diverse algebras underlying these integrable models are also
reducible from the ancestor algebra or its quantum-deformation
\cite{kunduPRL99}.  There is a separate class of models with long-range
 interactions \cite{Cal,Mos,Sud}, which
although are solvable
 quantum many body systems,  exhibit  different properties than those listed
above and  will not be  discussed here.
  The ancestor model scheme, though a significant achievement in unifying
and generating integrable models, seems to be also an apparent
disappointment, since it looks like a no-go theorem, allowing no
construction of new
 integrable  models beyond the known ancestor model. Moreover, since
the ancestor model and hence all    QI models as its descendants, are 
 defined  in 1D, it   apparently   excludes any construction  of 
  integrable  quantum models in higher space dimensions. 2D Kitaev models
\cite{kitaev1,kitaev2}, belonging to a different class,
  are possibly the only exception.

Therefore for a breakthrough, we look for new ideas and observe, that 
 the rational  ancestor Lax matrix   depends on the spectral
parameter $\la $ only linearly, while its q-deformation depends
  on its trigonometric functions  \cite{korbook,kunduPRL99}. 
  Consequently, all quantum
Lax matrices of  known integrable models, since realized from the ancestor
model,   depend also linearly (for the rational class),  or trigonometrically on
 $\la $  (for  q-deformed  class).  For going beyond the
prescribed form of the ancestor model, we search for a Lax matrix with
higher {\it scaling} (or length) dimension linked to the integrable hierarchy 
and for introducing extra space dimension, 
  exploit the concept of multi-space-time \cite{suris,kunduarXiv12}
$x_n, t_n, \ n=1,2,3, \ldots  $ ,
 hidden in integrable systems.
For a concrete application we confine  to the  $n=2 $ space case and 
propose  an alternative Lax matrix  approach with
$\lambda ^2 $ dependence, 
  focusing  on  the NLS field  model as an  example.  It
is quite surprising, that though such higher order  Lax matrices (with higher
order poles) are
  well known in the context of classical integrable systems, 
they have never been  used, as far as we know,
 in the construction of quantum
models. 
Note that, taking   $x=x_1,\ y=x_2,\ t=t_3 $ in the NLS hierarchy, would result to
the inclusion of an extra space-dimension $y$, apart from $x$  and the construction of    a novel
  quasi-$(2+1)$ dimensional NLS quantum field model,
 involving  a scalar  field $q(x,y,t) $ and its conjugate $q^\dagger $.
 For confirming  the complete integrability of the model, one needs to 
 show the mutual commutativity of all its conserved operators, which is
 guaranteed when  the associated Lax matrix satisfies the QYBE. 
However, this task for the present  Lax matrix turns  out to be the most difficult one, since the
commutation relations (CR) for the basic fields, known for the existing QI
models fail here, due to  significantly  different structure of our Lax
matrix and its higher $\lambda $ dependence. Moreover,  we  can no longer  seek  the
guidance of the ancestor algebra \cite{kunduPRL99}, since we have gone
beyond the known ancestor model.  Fortunately, we could  discover
intriguing algebraic relations for our basic quantum fields, which solve  the
required QYBE with the known rational  $R$-matrix.  Since the QYBE not
only proves the integrability of a quantum model, but also gives the CR
between the generator of  conserved operators and the generalized {
creation} operator, we can go ahead with the application of the algebraic BA
to our quasi-2D quantum field model and solve  exactly the EVP for all
its conserved operators including the Hamiltonian.  Many particle scattering
and  bound states differ considerably from the known result for the 1D NLS
model. The  bound states, 
corresponding to a  complex  solution for the particle
momentum, are found to  exhibit unusual properties with a variable  stability
region, dependent on the  particle number,   coupling constant and the
average particle momentum.

\section{ Quantum integrable models as descendants} 
QI models are  associated with a discretized  quantum Lax matrix $U^j(\la ) $, the  
 operator elements of which, for ensuring  the integrability of the model, must  satisfy
certain algebraic relations, which are  expressed 
 in a compact matrix form through  the   
  QYBE
\be    R(\la-\mu) \ U^j(\la ) \otimes U^j(\mu ) =
U^j(\mu ) \otimes U^j(\la ) R(\la-\mu),\ll{qybe}\ee
at each lattice site $j =1,2, \ldots N, $ 
  together with an {\it ultralocality} condition 
 \be  [U^j(\la )\otimes  U^k(\mu)]=0, \  \ j\neq k . \ll{Ulocal}\ee



Individual Lax matrices, each  representing a particular integrable model,
differ substantially  in their structure, content, nature of the basic
fields and  underlying algebras, whereas the quantum $R$-matrix, appearing in the QYBE
as structure constants,   remains the  same for all models belonging to the same
class and therefore can be of only three types: rational, trigonometric and
elliptic. However,  in spite of widely different
   Lax matrices linked to the rich variety of known QI models, they  are 
 in fact   realizable from a single rational ancestor Lax matrix
 or its q-deformed trigonometric form \cite{kunduPRL99}.
 We will not be concerned here with the elliptic models, which are anyway few in
number.  
The rational ancestor Lax matrix taken in the form
 \be
  U_{rAnc}{(\la)} = \left( \begin{array}{c}
 {c_1} (\la + {s^3})+ {c_2}, \ \ \quad 
  s^-   \\ 
    \quad  
s^+ ,   \quad \ \ 
c_3 (\la - {s^3})- {c_4}
          \end{array}   \right),
 \ll{LK} \ee
 satisfies the QYBE with the well known
rational $R$-matrix \cite{fadrev}, due to its underlying
 generalized spin algebra 
 \be [  s^-, s^+]= 2m^+ s^3+m^-, \ 
 [s^3, s^\pm] = \pm s_j^\pm \ll{ancAlg}
\ee where $m^+=c_1c_3, m^-= c_2c_3+c_1c_4$, with $c_j $s  as Casimir
operators or   
constant parameters admitting
zero values, and is capable of generating the known   quantum  integrable models of
the rational
class.
The rational  quantum
$R(\la-\mu) $ matrix in its $4 \times 4 $ matrix representation may be   defined
through its nontrivial elements  as \bea R^{11}_{11}=R^{22}_{22 }\equiv  a (\la-\mu) =\la-\mu
+i\alpha,   & &     \nonumber \\ R^{12}_{21}=R^{21}_{12}  \equiv  b (\la-\mu) =\la-\mu  ,
   \ \  R^{11}_{22}=R^{22}_{11}  \equiv   c=i\alpha, 
 & &  \ll{Rrat} \eea 
while  the trigonometric
case has q-deformed elements:  $a={\rm sinh }(\la-\mu
+i\alpha), \  b={\rm sinh }(\la-\mu), \  c={\rm sinh }(i\alpha).\ $ 
  The representative Lax matrices of known QI models of the
rational class can be recovered from the rational ancestor model
(\ref{LK}). We present below  a few of  such examples to illuminate the
situation. 
A general form for the Lax operators, which can be realized through a
bosonic representation from (\ref{LK}) was proposed earlier \cite{korbook} (Ch.
VIII.4).   
\subsection{Generation of rational models}

 {\it ${\rm xxx}$-spin chain} \cite{xyz}: The  Lax matrix may be reduced  from the ancestor
matrix (\ref{LK}) at 
 \ \   ${c_1}=c_3=1,\  \
 {c_2=c_4}=0,$ giving $m^+=1, m^-=0 ,$  which transforms  ancestor algebra (\ref{ancAlg})   
 to the spin algebra for 
   Pauli matrices.
 
 {\it  Lattice NLS model} \cite{korbook}: The Lax matrix may be  obtained from  (\ref{LK}) 
 at the  above parameter values,
by  mapping   spin operators   through
 the Holstein-Primakov transformation  to the  bosonic
operators: \ $[q_j, \ q^\dagger_k]=\delta_{jk}, $.

 {\it NLS field model} \cite{fadrev}:  The  Lax matrix may be recovered    from its lattice version
 at the field limit, 
giving the simple  familiar form
 \begin{eqnarray}
U_{(1)}= i\left(\begin{array}{cc} \la &   q \\ q^\dagger  & - \la
\end{array}\right)
\ll{Lnls} \end{eqnarray}
 with  bosonic field CR: 
$[q(x), \ q^\dagger(x')]=\delta (x-x{'}).  $

{\it Toda chain} \cite{toda}: \ Lax matrix may be    obtained  from  (\ref{LK}) at the parameter choice 
  ${c_1}=1, \  \  
{c_2=c_3=c_4}=0, $ resulting both  $m^\pm=0,$   with  generators of 
 the reduced   algebra  realized through canonical variables
$ [q_j,  p_k]=\delta_{jk}$.

 The rest of the QI models
of the rational class, like ${\rm xxx}$-Gaudin chain,  tJ and Hubbard
model etc.  can also  be covered by  rational ancestor model (\ref{LK}), employing
limiting procedures, higher rank representations, fermionic realizations
etc., details of which we skip.
\subsection{Trigonometric models}
Similarly  QI models belonging to the trigonometric class, e.g.
${\rm xxz}$ spin chain, relativistic Toda chain, sine-Gordon  model,  Liuoville model,
 derivative NLS model etc., are   derivable  from their
representative Lax matrices, which   in turn can      be  generated 
  from a single trigonometric ancestor Lax matrix. This ancestor  matrix is a 
  q-deformation of (\ref{LK}) and  satisfies the QYBE
with the trigonometric $R$-matrix, due to its underlying generalized
quantum group algebra.
 The  details, which  we omit here, can be found  in  
 \cite{kunduPRL99}.

\section{  Novel  quasi-2D  NLS model}    
Since the known quantum Lax matrices (including (\ref{Lnls})), as discussed above, inherit their
properties from the ancestor models, all of them depend on the spectral
parameter $\la $ linearly  (for  rational  models) as in (\ref{LK}), 
or on ${\rm sin}\la , {\rm cos} \la $ functions 
 (for  trigonometric models) and are defined in one-dimensional space.
A prominent example of  rational models is the $(1+1) $-dimensional NLS
field model   associated with the Lax matrix (\ref{Lnls}).
Therefore for going beyond the 
known models and introducing extra dimensions,
 we look 
into the background concept of multi-space-time dimension $\{ x_n\}, \{t_n\}
, \ n=1,2, \ldots N, $ hidden in the theory of integrable systems
\cite{suris,kunduarXiv12}. In this formulation of multi-dimension one can
define multiple Lax equations of the form
 \begin{eqnarray}
T_{ x_n}= U_{(n)} T, \ \  T_{ t_n}= V_{(n)} T, \   n=1,2, \ldots N,  
\ll{Txt} \end{eqnarray}
(here and  what follows we denote partial derivatives as subscripts, as
a short-hand notation) 
 where $T=T(\lambda,q) $ is the monodromy matrix dependent on the field
$q=q(\{ x_n\}, \{t_n\}), $  defined in multi-space-time and the generators
of the infinitesimal  space-time translation $ U_{(n)}(\lambda),  V_{(n)}(\lambda) $ are the
space and time Lax operators, respectively. However, since  the set of Lax
equations (\ref{Txt}) is a overdetermined system, its compatibility
conditions (equality of mixed derivatives) would lead to the pairing between 
any two Lax matrices:  $ (U_{(n)},  V_{(m)}), (U_{(n)},  U_{(m)}), (V_{(n)},  V_{(m)})
$,  due to
symmetry among the variable.  Consequently, this would lead to the flatness condition among each Lax
pair, generating a series of classically  integrable hierarchal
equations as
 \begin{eqnarray}
\partial_{ t_m}U_{(n)}-\partial_{ x_n} V_{(m)}+ [ U_{(n)} V_{(m)} ]=0,
\nonumber \\ 
\partial_{ x_m}U_{(n)}-\partial_{ x_n} U_{(m)}+ [ U_{(n)} U_{(m)} ]=0, 
 \ \   n,m=1,2, \ldots N, 
\ll{flatness} \end{eqnarray}
etc. and similarly with other pairs.
  A possible reduction $U_n=-V_n $ may be
introduced due to the simplified 
condition $\partial_{ t_n}U_{(n)}-\partial_{ x_n}
V_{(n)}=0. $    
The   hierarchal
equations (\ref {flatness}) represent integrable systems in $(1+1)
$-dimensions,
 $2 $-dimensions or in
quasi-higher dimensions.  

We intend to use this concept of   the
 hierarchy of multiple space-times, embedded  in  integrable
systems, for constructing quantum integrable models in quasi-$(2+1) $
dimensions, restricting to the case  $q(x_1,x_2,t_3), $
 involving two space  and one time
Lax operators $(  U_{(1)}, U_{(2)},  V_{(3)}), $ linked to space  $x=x_1, y=x_2
$ and time $ t=t_3 $ variables.    
 \subsection{Alternative Lax matrix }
For a  concrete application, 
we consider the  NLS family  of field models, which  
belongs to the rational class and choose   
our quantum Lax matrix as its next  hierarchy: 
 \begin{eqnarray}
 U_{(2)}(\lambda)= -i\left(\begin{array}{cc}  2 \la^2  -q^\dagger q &   
2 \la q-iq_x \\ 
-2 \la q^\dagger-iq_x^\dagger  & - 
 2 \la^2  +q^\dagger q
\end{array}\right)
\ll{V2} \end{eqnarray}
It is interesting to compare the structure of  Lax  matrix (\ref{V2}) with 
that of the well known NLS model (\ref{Lnls}), to
 note the  crucial differences, that the  matrix elements of (\ref{V2})
 depend on the spectral parameter up to  $ \la^2 $ (double pole) and involve 
     field operators
$q, q^\dagger, q_x, q^\dagger_x , $ defined in quasi-$(2+1)$ dimensions: $
(x,y,t)$. 
It needs to be mentioned, that  such higher order  Lax
matrices like    (\ref{V2}) appearing in the integrable hierarchy are
usually taken  to be independent entries,  not constructed solely out of
lower order Lax matrices, though there are formalisms to connect them in an
involved  way using classical $r$-matrix \cite{FadTakhHam}. However no
quantum extension of this method is available and it is also not clear
whether the general L-operators of \cite{korbook,kunduPRL99} can be used for
this purpose.  
Therefore, leaving aside the question about the possibility
of  constructing  (\ref{V2}) from more fundamental Lax operators,
we start directly with 
this  higher order Lax operator   for constructing our quantum
model.   

We emphasize again that, in spite of the familiarity of    Lax
matrix    (\ref{V2})  in   classical
integrable hierarchy, such higher-pole Lax operators   
    have   been ignored so far   in
the context of quantum integrable models.
Note that, in the hierarchal equations  we consider the pairing 
$(U_{(2)},  V_{(3)}) $ with $x=x_1, y=x_2, t=t_3, $ for   constructing
quasi-$(2+1) $ dimensional model, which can also
 be reached by a combination of
the Lax pairs $(U_{(1)},  U_{(2)}) $ in $2$-dimensions with $x=x_1, y=x_2$
and $(U_{(1)},  V_{(3)}) $ in $(1+1)$-dimensions with $x=x_1, t=t_3$.    
 Note that, the $x$-shift space Lax operator here can be given by $U_{(1)}$ in
(\ref{Lnls}),
 associated with  the space Lax operator of the standard
NLS model, while the
  $y$-shift space Lax operator is given   by $U_{(2)}$ in
(\ref{V2}) and the  $t$-shift time  Lax operator  by $V_{(3)}$,
representing a higher order $\lambda^3 $ dependence (cubic pole) form, which
we do not reproduce here. However, for constructing our quantum model and
exactly solving it  through algebraic Bethe ansatz we would  need only
the quantum Lax operator     $ U_{(2)}.$

We  introduce here    the notion  of scaling or length dimension,
which is an useful concept in analyzing higher order Lax operators in
multi-space-time dimensions. One defines a scaling dimension 
$[L^{-1}]=1$  for length $L$. Therefore from (\ref{Txt}) we get
$[U_{(n)}]=n, $ since $[\partial_{x_n}]=n $
(similarly for $V_{(n)} $ ) and consequently $[U_{(1)}]=1$, since  each term in  
(\ref{Lnls})   has scaling dimension $[\lambda
]=[q]=1 $. Similarly, $[U_{(2)}]=2$, since
in
(\ref{V2}) 
 $[\lambda^2
]=[|q|^2]=[q_x]=2 ,$ and  $[V_{(3)}]=3,$ etc.

\subsection{Quantum integrability through Yang-Baxter equation }
In dealing with quantum field models one has to lattice regularize
the Lax operators first to avoid short-distance singularities \cite{fadrev}. Therefore,
our  intention is to show, that the discretized  Lax matrix along the
$y$-direction:
  $U^j=I+\De U_{(2)}(\lambda, q_j),$ where $q_j= q (x,y=j,t) , $
  with   lattice constant $ \De \to 0$, 
does satisfy  the QYBE (\ref{qybe}) with the rational $R$-matrix (\ref{Rrat}).
 However, this becomes a
highly involved  problem, since  due to  more complicated
structure of the present Lax matrix (\ref{V2}), 
{\it ten} out of  total 16 relations of the
$4\times 4$  matrix QYBE  remain nontrivial, all of which are to be
satisfied with a suitable field CR.
  Compare this situation 
with the  known  1D NLS case  \cite{fadrev,sklyaninNLS},
  where due to much simpler form of the  Lax
matrix $ U_{(1)} $  (\ref{Lnls}), only {\it two} nontrivial
  relations in the  QYBE survive, which can be solved     
successfully     using   the  bosonic field CR.
 However, we realize  that,
 no   algebraic relations, including   the  bosonic  CR, appearing in 
 the existing    integrable 
models would work here, since the choice of  $U_{(2)} $ has taken us  beyond the
scope of the known
ancestor models and the associated  algebras. Moreover, the CRs for the
field now have  to be  sought for along the extra direction $y$, that has
been included in the system.  Therefore we  look 
for some  innovative algebraic  relations    for the  basic quantum fields
to be consistent with the QYBE, linked to the present Lax matrix 
(\ref{V2}). Fortunately, we  find
 a new set of such relations  for our quasi-2D  fields as 

\bea 
[ q(x,y,t), \ {q}^\dagger_x(x,y^{'},t) ]=-2i \alpha \  \delta(y-y{'}), \ 
[ q_x(x,y,t), \ {q}^\dagger(x,y^{'},t) ]= 2i  \alpha \delta(y-y{'}), 
  \label{CRx} \eea
\be
[ q(x,y,t), \ {q}^\dagger(x,y^{'},t) ]= 0,
 \label{CR0}
\ee
(or their discretized version (\ref{CRj}))
together with  their hermitian conjugates. 
Note that CRs (\ref{CRx}) (or (\ref{CRj})), 
exhibiting  an asymmetry in space variables are fundamentally  new relations,
different from  known relations like canonical, bosonic etc.  
It may be observed that, the form of CR  (\ref{CRx}) may be linked
to the quadratic {\it space-velocity} term $:q^\dagger_xq_x:$ appearing in Hamiltonian
(\ref{Hx}). 
Application of these  algebraic relations (\ref{CRx},\ref{CR0})
   satisfies   miraculously all ten nontrivial  
equations appearing  in the  QYBE, involving  the discretized Lax matrix $U^j(\De ) $
, up to
order $O(\De ) $ (see (\ref{App_Deqybe}-\ref
{App_exactqybe}
) in { Appendix} for details). This is however  enough for proving the integrability of 
the   field models,   obtained  at  
the limit $\De \to 0 .$ It is remarkable, that in spite of the presence of
a $x$-derivative  term,  new CRs (\ref{CRx})  satisfy  the necessary
ultralocality condition (\ref{Ulocal}). This is because not $x$ but  
$y$ is  the  relevant space direction  here, where the fields  commute    at space separated
points  along   $ j \to y$, 
    reflecting the
quasi-2D nature of our model.  

Therefore, since due to 
(\ref{CRx},\ref{CR0}) (or its discretized version (\ref{CRj}))
, 
 the lattice regularized quantum Lax operator $U^j(\la ) $  constructed
from  (\ref{V2}) satisfies
the QYBE  (\ref{qybe}) for the rational $R$-matrix,
 together with  the ultralocality condition (\ref{Ulocal}),
 the transition matrix for our model,  defined    for $N$-lattice sites:
 $T(\la)=\prod_{j=1}^N U^j(\la ) , $  must also satisfy the QYBE
 \cite{fadrev}
\be    R(\la-\mu) \ T(\la ) \otimes T(\mu ) =
T(\mu ) \otimes T(\la ) R(\la-\mu),  
\ T(\la ) = \left( \begin{array}{c}
 A(\la ), \  \ \quad 
  B(\la )  \\ 
  B^\dagger(\la ) ,   \quad \ \ 
A^\dagger(\la )
          \end{array}   \right),
\ll{qybeg}\ee
 with the
same $R(\la-\mu)$-matrix.  This happens due to the coproduct property of the underlying
Hopf algebra \cite{chari}, which keeps an algebra invariant under its tensor
product.  This global QYBE (\ref{qybeg}) serves two important purposes.  First, it proves
the quantum integrability of the model by showing the mutual commutativity
of all conserved operators.  Second, it derives the commutation relations
between the operator elements of $T(\la )$, which are used for the exact
algebraic Bethe ansatz  solution of the EVP.

In more details: multiplying  QYBE 
(\ref{qybeg}) from left by $R^{-1} $,  taking the trace from both  sides  and
 using the property of cyclic rotation of matrices under
the trace,  
one can show   that $ \tau (\la) = {\rm trace} \ T(\la) $  commutes:
 $[\tau (\la),\tau
(\mu) ]=0. $  This in turn   leads to the Liuoville integrability condition:
 $[C_j,C_k]=0, \ j,k=1,2, \cdots $, 
  since the   conserved set of local operators are generated from $\ln
 \tau(\la)=\sum_j C_j \lambda ^{- j}, $ through expansion in the spectral
parameter $\la  $.  Following this construction and
 exploiting the explicit form of the Lax matrix (\ref{V2}),
  we can derive, in principle, 
 all conserved operators $C_j, \ j=1,2,\ldots $ for our model.
 Skipping  the details,  which can be found 
 for the classical case in \cite{kunduarXiv12}, we present here  only the
 $x$-shift  
Hamiltonian as $ H_{(x)} \equiv  { C}_2 $:
\bea  H_{(x)} 
 = \int dy : (
iq^\dagger q_{y}+q^\dagger_{x}q_x +{q^\dagger}^2q^2):
  \ll{Hx}\eea
and the $t$-shift  
Hamiltonian as $ H\equiv  { C}_4 $:
\bea  H &=& \int dy : (
iq^\dagger_{x}q_{xy}+q^\dagger_{y}q_y+
i    (q^\dagger q)(q^\dagger q_{y}- q^\dagger_yq)\nonumber \\
 &-&2 (q^\dagger q) \ q^\dagger_xq_{x}+ {q^\dagger}^2q^2_{x}+
{q_x^\dagger}^2q^2
 ):, \ll{C4}\eea
which we take as  our model Hamiltonian, we are interested in. 
Notice the    quasi $(2+1) $ dimensional nature of   Hamiltonian
 (\ref{C4}), since though it involves both $x$ and $y$ derivatives of the field 
: $ q_{x}(x,y,t)$ and $q_{y}(x,y,t), $
the volume  integral is taken  only along $y$.
Asymmetry in the appearance of  space derivatives
is also explicit. However, at the same time an operator with double
space volume integral: ${\rm H} =\frac 1 L\int ^L_{-L}dx \  H $ in a strip of $x \in [-L, L] $ is
 also conserved in time, since $
\partial_t H=0 . $  
\section{ Algebraic Bethe ansatz  for the eigenvalue problem}
 Since $T(\la)$
satisfies the QYBE (\ref{qybeg}) with the rational $R$-matrix, we can follow the procedure
for the algebraic BA, close to the formulation for the 1D quantum
NLS model \cite{fadrev,sklyaninNLS}.  As we have discussed above, $ \tau( \la )={\rm
trace}  T(\la)=
 A(\la ) + A^\dagger (\la )$ is linked to the generator of the conserved
operators $C_j, \ j=1,2, \ldots $, including the Hamiltonian (\ref{C4}). 
The off-diagonal elements  of 
 $T(\la )$ (\ref{qybeg}): $ B(\la)$ and $B^\dagger(\la)$,  on the other hand,
 can be
considered as   generalized {\it   creation} and {\it   annihilation}
operators, respectively.
For solving  the EVP for all conserved operators:
$C_j|M>=c^{(M)}_ {j}|M>, \ j=1,2, \ldots $  simultaneously,
we  construct exact M-particle Bethe state  $|M>= 
B(\mu_1)B(\mu_2)\cdots B(\mu_M)|0>, $  on a  pseudo-vacuum $|0> $ with the
property \  \ 
 $B^\dagger (\mu_j)|0> =0, \  A(\la)|0>= a_0(\la) |0>, $ \  and aim to solve 
the EVP:\ \  $\tau (\la )|M>= \Lambda_M (\la, \mu_1, \mu_2,\ldots,  \mu_M )
|M>, \ \  $
with exact eigenvalues   $\ln \Lambda_M (\la, \{ \mu_a \})=\sum_j c^{(M)}_j ( \{ \mu_a \})
 \lambda ^{- j}$.

\subsection{Exact solution for  quasi 2D quantum field  model}
For obtaining the final  result for our quantum NLS field model on infinite
space interval we follow the formulation of \cite{fadrev} for the 1D NLS
field model, though adopted here for the transverse dimension $y$ and higher
order Lax operator (\ref{V2}).  
We  switch over to the field limit: $\De \to 0 $ with total lattice site
  $N \to \infty $ and then take the interval $L=N \De \to \infty, $ 
assuming vanishing of the  field $q_{j }
 \to 0, \ \mbox {at  } j \to \infty $,  compatible with  the natural
boundary condition of having the
vacuum state  at  space   infinities, yielding the asymptotic Lax matrix
 $U^{j}(\la )|_{j\to \infty}=U_0(\la )=2i \la ^2 \sigma^3.$ 
Therefore, we have to  shift over  to the field transition  matrix defined as
\be T_f(\la )=U_0^{-N} \ T(\la )  \  U_0^{-N}, \ \ \  N \to \infty , \ll{Tf} \ee
and for  further construction  introduce 
 $\ \ V(\la,\mu) \equiv U_0(\la )\otimes U_0(\mu ) , \ $
 $W(\la,\mu)=(U^j(\la )\otimes
U^j(\mu ))_{j \to \infty} .$
  We may   check from the QYBE (\ref{qybe}) that $W $ satisfies the relation
   $ \ R(\la - \mu) W(\la, \mu)= W(\mu, \la) R(\la - \mu),  \ $
using which  we can derive  from  QYBE (\ref{qybeg}),
 that the  field transition  matrix (\ref{Tf}) also satisfies the QYBE
\be    R_0(\la,\mu) \ T_f(\la ) \otimes T_f(\mu ) =
T_f(\mu ) \otimes T_f(\la ) R_0(\la,\mu),
\ll{qybegf}\ee
but  with a  transformed $R $-matrix: 
\be R_0=S(\mu,\la) R(\la -\mu) S(\la, \mu), \ 
S(\la, \mu)=W^{-N}(\la, \mu)V^N(\la, \mu),   \  N \to \infty , \ll{R0} \ee  where
 $R(\la -\mu)$ is the original rational $R -$matrix (\ref {Rrat}) 
(see \cite{fadrev}  for similar details on  1D NLS model).

Based on  the above  formulation and using the field operator products:
$q_jq^\dagger_{j,x}=- 2i\frac{ \alpha} \De, \ q^\dagger_{j,x} q_j=0,   $ at $j \to
\infty , $   compatible with the field CR, 
we can calculate  explicitly  the relevant objects needed for our field model. In
particular,
the  central $2 \times 2  $  block $W_c $ for 
matrix $ W $ turns out to be 
\bea W_c(\la, \mu)= I+\De \ { M}(\la, \mu)
\left(\begin{array}{cc}   (\la - \mu)   &   
0 \\ 
-2 \alpha & - 
 (\la - \mu)
\end{array}\right), 
 \ll{W} \eea
with an intriguing factorization of its spectral dependence  by a
prefactor $ { M}(\la, \mu)=2 (\la + \mu) , $ which is  the key reason behind the success
of 
the  exact algebraic Bethe ansatz solution for our  field model, in spite of
the  more complicated form of  its Lax operator.
Note, that since our model shares  the same    rational $R$-matrix with   the known  NLS
case (though having different    
Lax operators), the present result coincides in part   
 with  that of the 1D  NLS model    \cite{fadrev,sklyaninNLS},
 though only formally. 
On the other hand, the transformed $R_0$ matrix, relevant for the
field model, depends  on the corresponding
asymptotic  Lax matrix and its product through matrix $S(\la,\mu) $.
Therefore, since  Lax matrix (\ref{V2}) for our model
is  more complicated, compared to (\ref{Lnls}) for  the
1D NLS model,     
our final result  shows intriguing differences from the known NLS result, which we
 highlight below. 
 
For constructing $R_0$  using  definition (\ref{R0}),
we have to construction first matrix $S(\la,\mu) $, taking proper limit of 
$W^{-N} $ at $ L  \to \infty $  using  (\ref{W}) .
Through some algebraic manipulations, which are skipped here, we finally arrive at the field limit
 to a  simple  expression for  $R_0$ matrix, expressed through its
nontrivial elements as
 \bea R^{11}_{11 }=R^{22}_{22 }=  a (\la-\mu) 
,   R^{12}_{21}=    b (\la-\mu) , \     R^{11}_{22}=R^{22}_{11}  =0,
   & & \nonumber \\ 
  R^{21}_{12} = b (\la-\mu)  - \frac
{\alpha ^2} {\la-\mu } + \frac
{\alpha ^2 \pi} {M(\la , \mu) } \delta(\la - \mu), & &     
   \ll{R0jk} \eea 
where ${ M}(\la, \mu)=
2 (\la + \mu) $, the terms $ a(\la-\mu), b(\la-\mu)$ are   as in  (\ref{Rrat})
 and the $\delta (\la -\mu) $ term
vanishes at $\la \neq \mu  $.  
It is interesting to compare $R_0 $-matrix (\ref{R0jk}), 
relevant for the field models,  with the original $R$-matrix
(\ref{Rrat}). 
Now from  QYBE (\ref{qybegf}) linked to field models, we can derive 
   for our model  the
required CR between the operator elements
 of $T_f $, using   
$R_0$ matrix (\ref{R0jk}). 
In particular, we get the commutation relation 
 \be A(\la) B(\mu_j)=(f_j(\la -\mu_j)+  \frac
{\alpha ^2 \pi} {{ M}(\la, \mu_j) } \delta(\la -\mu_j))
 B(\mu_j ) A(\la),  \ll{AB} \ee 
 where $ \ f_j= \frac 
{\la -\mu_j - i \alpha } {\la -\mu_j
 }. $  Note that the 
 prefactor  ${ M}^{-1}(\la, \mu_j)= \frac 1 {
2 (\la + \mu _j)} $ appearing in the above CR  bears the imprint of
the  $\la ^2 $ dependence of our Lax matrix and is absent in such relations
in the standard NLS model. 
At $\la  \neq \mu_j $ however,  when the singular term
 vanishes,  the relation 
  coincides formally with the known NLS case.
Using this result and the property of the vacuum state $|0> , $ 
we obtain  the   exact  EVP for $$A(\la)|M> =  F_M |M> , \ \mbox{ as }
 \ F_M = \prod_j^M f_j(\la
-\mu_j)$$ and hence for $\tau(\la ) $,  which gives finally the exact  eigenvalues for all
conserved operators. For our model Hamiltonian $H=C_4 $,
 we obtain 
 the exact energy spectrum  $E_M=\sum_j^M \mu_j^4 $, for the M-particle
scattering state, which clearly differs from that of the known NLS model
\cite{fadrev,korbook}, though bearing formal similarity with the next NLS
hierarchy.

  It is remarkable, that in spite of the highly
nonlinear field interactions present in the Hamiltonian (\ref{C4}), the scattering
spectrum shows no coupling between individual quasi-particles, mimicking a
free-particle like scenario.
On the other hand,  the bound-state or the quantum {\it soliton} state, which
is  obtained
for the complex   {\it string} solution for the particle  momentum:
  $ \mu^{(s)} _ j =\mu _0 +  i \frac \alpha 2 ( (M+1)-j), $
where $\mu _0 $ is the average particle  momentum 
 and  $ \alpha $ is the coupling constant, induces  mutual interaction
   between the particles.  The corresponding energy spectrum may be given by
   \be E_{M}^{(bound)}= Re[\sum_j^M
({\mu^{(s)}_j})^4]= M\mu^4 _0+E_b(\mu_0, \alpha, M), \ll{EMb} \ee
 where $E_b $ is the
binding energy of the $M > 1$-particle bound-state.  Recall, that  a bound-state
becomes stable, when its energy is lower than the sum of the
individual free-particle energies, which in turn is ensured by the {
negative}
values of the binding energy: $E_b $.  More negative binding energy
indicates more stable bound-states.  For the known NLS model the binding
energy \cite{fadrev,sklyaninNLS} $ \ E_b^{nls}(\alpha,M)=-\frac
{\alpha^2} {12}M(M^2-1) \ $
is independent of  $\mu_0 $ 
and strictly negative,
which  makes the corresponding bound-states always stable with the stability increasing
as  the particle number $M$ and the coupling constant
$\alpha $   increase.

  However,   
for  the  present quasi-2D NLS model,  the picture   differs  
  significantly, producing a fascinating bound-state  spectrum
 with intricate stability region. Note, that in the present case the   binding
energy  
 \bea 
E_b(\mu_0, \alpha, M)= E^+ +E^-,    E^-= - \frac
{\alpha^2\mu_0^2 } {2}M[ (M^2-1) & &  \nonumber \\
  E^+= \frac{\alpha^4} {16} M[(\frac{1}
{5}M^4 +1 -
 \frac{2} {3}M (M+ \frac{4} {5}))],  & &   
  \ll{Eb}\eea
contains both { negative}  and { positive}  terms (due to the simple
mathematical  fact, that in the   expression \ \  $\sum_j^M
({\mu^{(s)}_j})^4$ \  the  real term with  $i^2=-1$ gives negative, while
 $i^4=+1 $  strictly positive contribution). Therefore,  binding energy  (\ref{Eb}) 
may  take { negative} as well as  { positive} values,  depending on the
parameters 
$\alpha,$  $ \mu_0$
 and  $M$, defining a variable  domain for the existence of
stable bound states (see Fig.1). Note that the term $E^-=\frac {\mu_0^2}
6 E_b^{nls}, $ proportional to that of the known NLS model,  stabilizes the
bound-state, while the counter term $E^+$ has a destabilizing effect.
\begin{figure}[!h]
\centering
{
 \includegraphics[width=6 cm, angle=0]{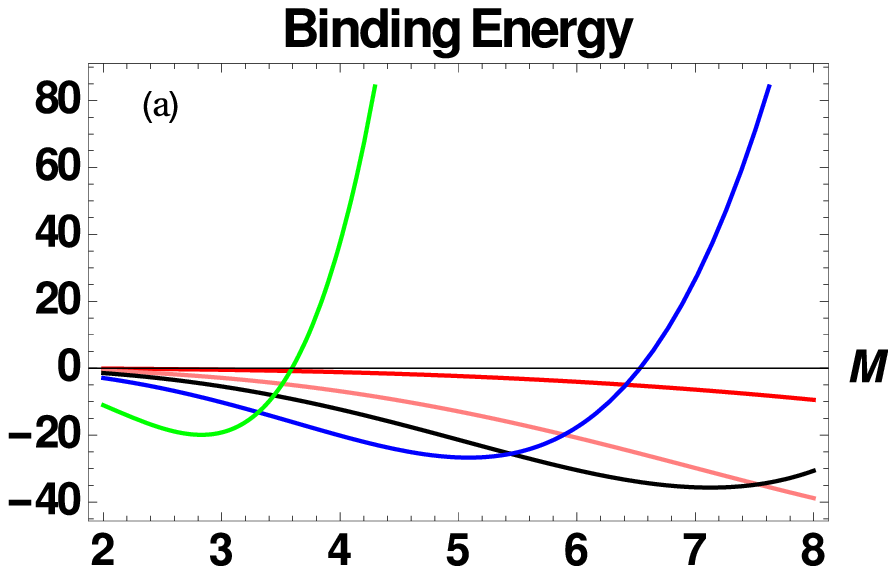} 
\vskip .8cm

\includegraphics[width=6 cm, angle=0]{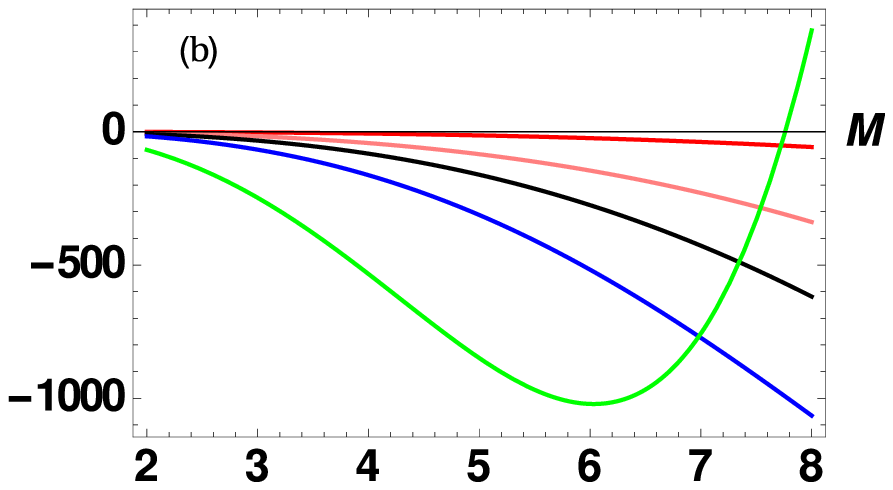} 
}
 \caption
 { Binding energy   for the quasi-2D NLS    model. Figure
 shows $E_b $ (\ref{Eb})
 with increasing particle number  $M$ for
different values of the coupling constant $\alpha$ =0.1 (red), 0.5 (pink), 0.7
(black), 1.0 (blue), 2.0 (green),
 with   parameter  $\mu_0 $  fixed at  (a)  $1.0$ and (b) $2.4 $.
}
\end{figure}
The graphs  show clearly  that   the stability region of the bound-state
 for our model
shrinks with the increase of the  particle number $M$  as well as with the 
coupling constant $\alpha $, which is rather
anti-intuitive, since for the known NLS model,as seen from $E_b^{nls} $, 
 the  bound-state  stability 
always increases with increasing $M$ and $\alpha$.
On the other hand,  one can enlarge the 
stability domain in our model by increasing the particle momentum $\mu_0 $,
 as evident by comparing the figures (a) and (b). This feature however
 can not be matched  with the known result  of the 1D NLS model, since its
binding-energy is independent of $\mu_0.  $ 
 This shows  the intricate nature  of the bound-state configuration for our
model in comparison with known result of the 1D NLS model.  

\section{ Conclusion and outlook}  Going beyond the
known form of the existing  integrable quantum models in 1D,
we propose an alternative higher order Lax matrix approach,
exploiting the concept of multi-space-time dimension  hidden in integrable systems, and
apply it
 for constructing and solving a novel quasi-2D quantum NLS
field model.   The key to our success in proving the crucial   quantum Yang-Baxter
equation, which guarantees the quantum integrability of the model, is the
discovery of a new type of field operator algebra,
 not covered by the existing   rules.

 The known 1D quantum integrable models satisfying QYBE with rational
R-matrix may be realized from an ancestor   Lax matrix   
   associated with a 
spin-like algebra,  reducible to  conventional  spin, bosonic,
or the  canonical algebra, related to the existing models.
 A bosonic realization of  this matrix
 was proposed as    a general L-operator in   Ch. VIII.4 of [3], followed by a
theorem, stating that the same L-operator can construct a  monodromy matrix
with arbitrary rational function. However this theorem, claiming only
 a sufficient but not a necessary condition,  does
 not rule out the possibility of  alternative L-operators, an example of
which is provided by the present Lax matrix.

We stress that, the quantum
  Lax operator (\ref{V2}) with
 higher scaling dimension  
 and the associated  novel  commutation relations
(\ref{CRx}-\ref{CR0}) for the fields in  our
model are fundamentally different from those used 
 in nonrelativistic integrable quantum systems.  Unlike the
known L-operators belonging to the rational class,  (\ref{V2})
   with  $\lambda^2$ spectral
dependence and having   x-derivative of the field, 
 can not be realized straightforwardly from any linear
combination of the general ultralocal  L-operators 
proposed  earlier \cite{korbook,kunduPRL99}.
 Similarly,  
  the crucial  algebraic structure (\ref{CRx}) (with   x-derivative
term) is
  different from  known ultralocal algebras  and evidently  can not be
 generated by combining them. However, a possible construction of such
higher-order Lax operators as a nonlinear combination (like product) of
lower order Lax operators of  \cite{korbook,kunduPRL99} and obtaining the  
underlying novel algebras from the known ones
could  be taken up  as  a challenging future problem.

The dimensionality of the present model with its field $q(x,y,t) $ needs
 special focus. In one hand, the system 
shares  effectively one-dimensional  properties, since it is linked to the
1D  NLS hierarchy. This is also reflected in the energy spectrum of the
present model, which is similar to that of the higher Hamiltonian in the
known NLS hierarchy.
  On the other hand, the  model  Hamiltonian (\ref{C4}) contains
derivatives  of the field $q_x, q_y $ in  both $x$ and $y$ variables and
similarly  both these variables are involved  in commutators (\ref{CRx}-\ref{CR0})
as well as in the present Lax matrix, defining
the model in quasi
two-dimensional form. Moreover,   these 2D structures can not be reduced to
1D
by ignoring the dependence on the other variable.

 Due to quantum integrability of 
our quasi-2D NLS  model, the  eigenvalue problem can be solved  exactly for 
the commuting set of all its conserved operators,  with    
   intriguing result for the many particle scattering and  bound states.

It is worth adding that,  recently we have
constructed a novel quasi 2D quantum Landau-Lifshits model belonging also to
the rational class (to be  reported elsewhere).
It is  reasonable to assume therefore, that
 such quasi 2D quantum models generated by higher-order Lax operators, are
not limited only to the present NLS case, but constitute a novel family of
quantum integrable systems within the rational class.
 The present approach,   general enough for  applying   to other quasi
higher dimensional  quantum   models, could open up a new direction
 in the theory of quantum integrable
systems. It is a challenge to find  
 a possible q-deformation of the  algebra  found here, which 
 could lead to  a novel class of quantum algebra, 
while an exact lattice version of the present Lax matrix could   unravel 
 a higher-order  ancestor Lax operator for  generating a new family  of integrable  quantum
models.

 \section{Appendix}
In  QYBE (\ref {qybe}) with $R$-matrix (\ref{Rrat}) and  discretized version  $
U^j$
of the  quantum Lax matrix (\ref{V2}), out of total 16 matrix operator
relations,  except 4 diagonal and 2 extreme
off-diagonal terms,  all other  10   relations $Q^{ij}_{kl} $  stand nontrivial and
 their validity needs to be proved using   the CR,  discretized from
(\ref{CRx}-\ref{CR0}): 
  \be [q_j,q^\dagger_{j,x}]=-2i \frac
{\alpha} \De, \ \quad [ q_j,q^\dagger_{k}]=0
 \ll{CRj} \ee    
  and their conjugates.  
\subsection{\it QYBE relation for matrix elements}  
Using expressions for $a(\la -\mu), b(\la -\mu), c $ of (\ref{Rrat}) and   CR
(\ref{CRj}) we may check the validity of  \bea 
Q^{11}_{12} 
 =a \ {U^j}_{11}(\la ) {U^j}_{12}(\mu )- b \ {U^j}_{12}(\mu ){U^j}_{11}(\la
)   - c \  
{U^j}_{11}(\mu ){U^j}_{12}(\la )) \nonumber  \\ = 
 i\De (\la -\mu)\ q (-\De [q_j^\dagger, q_{j,x}]+2 c ) \ \   +O(\De ^2) =0,
\ll{App_Deqybe} \eea 
 upto order $O(\De ^2)$.
Similarly one proves the conjugate relations $Q^{11}_{21}, Q^{21}_{11}, Q^{12}_{11}$
and similar relations $Q^{22}_{12}, Q^{22}_{21},$ $ Q^{12}_{22}, Q^{21}_{22} .$

The validity of the  remaining two relations can also be  proved
 with the use of the same  CR (\ref{CRj}): 
\bea Q^{12}_{21} 
 =b \ [{U^j}_{12}(\la ), {U^j}_{21}(\mu )]  \ \   +c \  ({U^j}_{22}(\la ){U^j}_{11}(\mu
)\ \  -
{U^j}_{11} (\la ){U^j}_{22} (\mu ))  \ \nonumber \\  = 
2i\De ^2
  (\la -\mu) (\mu [ q_{j,x},q_j^\dagger]+\la [ q^\dagger_{j,x}, q_j]  ) \ \
 + 4i \De c (\mu^2-\la^2)
\   =0 , 
\ll{App_exactqybe}\eea
which is valid
 exactly in all orders of $\De $ and
similarly for the conjugate relation $Q^{21}_{12}$.
This proves thus the
validity of all QYBE relations for our quantum quasi-2D NLS field model, associated
with the higher Lax operator (\ref{V2}) and algebraic relations
(\ref{CRx}-\ref{CR0}), obtained at the limit   $\De \to 0 $ .
\vskip 1cm


\begin{thebibliography}{99}

\bibitem{fadrev}  L. D.
 Faddeev ,
 { \it Quantum completely integrable models in field theory},
Sov. Sc. Review, C1 (1980) 107
\bibitem{kulskly} P. Kulish and E. K.
 Sklyanin, {\it Quantum spectral transform method. Recent developments},
Lect. Notes in Phys. (ed. J. Hietarinta et al, Springer,Berlin, 1982) vol. 151
p. 61.
\bibitem{korbook} V. E.
 Korepin, N. M.
 Bogoliubov, A. G.
 Izergin, {\it QISM
and Correlation Functions}, (Cambridge Univ. Press , 1993)
\bibitem{baxter} R. Baxter,
 {\it Exactly solved models in statistical mechanics}
(Acad. Press, 1981)

\bibitem{MattisEncyl} D. C.
 Mattis, {\it The Many Body Problems}, (World
Sc., 1993)
\bibitem{bethe31} H. Bethe,{\it Theory of matal I.  Eigenvalues and
eigenfunctions of the linear atomic chain}  , Z. Phys. 71 (1931) 205
\bibitem{xyz}   L. A.
 Takhtajan  and L. D.
 Faddeev, {\it Quantum inverse scattering method and the Heisenberg XYZ
model} , Russian Math. Surveys 34
(1979) 11-68
\bibitem{xxz} P. P.
 Kulish and E. K.
 Sklyanin,{\it Quantum inverse scattering method and the Heisenberg
ferromagnet} , Phys. Lett.  70 A (1979) 461 
\bibitem{dbose} E. Lieb and W. Liniger, {\it Exact analysis of an
interacting Bose gas. I. The general solution and the ground state },  Phys. Rev. 130 (1963) 1605

\bibitem{d1bose}
A. G. 
Shnirman, B. A.
 Malomed and E. Ben-Jacob, {\it Nonperturbative studies of a quantum
higher-order nonlinear Schr\"odinger model using the Bethe ansatz },  Phys. Rev.  { A 50} (1994)  3453 

 \bibitem{dany}
 A. Kundu, {\it Exact solution of double-delta function Bose gas through interacting  anyon gas}
, Phys. Rev. Lett. 83 (1999) 1275
 
\bibitem {d1any}
  M. T.
 Batchelor, X.-W. Guan, A. Kundu, {\it D-anyons: one-dimensional anyons with competing $\delta$-function and
derivative $\delta$-function potentials}, J. Phys. (FTC) A 41 (2008) 352002  
%
\bibitem{sklyaninNLS}  E. K.
 Sklyanin, DAN SSSR, 244 (1979) 1337 


\bibitem{kunToda}
 A. Kundu, {\it Generation of a quantum integrable class of discrete-time or relativistic
 periodic Toda chains},  {  Phys. Lett. A}190 (l994) 79-84 

\bibitem{toda}  E. K.
 Sklyanin,{\it The Quantum Toda Chain},
Lect. Notes in Phys.  226 (1985) 196-233


\bibitem{tj}   F. Essler and V. E. Korepin,{\it Higher conservation laws and
algebraic Bethe ansätze for the supersymmetric t-J model }, Phys. Rev. B46 (1992) 9147
 
\bibitem{hubb} E. H.
 Lieb and F. Y.
 Wu, {\it Absence of Mott transition in an exact solution of the short-range
one-Band model in one Dimension}, Phys. Rev. Lett. 20 (1968) 1445
\bibitem{hubbS}B. S.
 Shastry, {\it Exact integrability of the one-dimensional Hubbard model}, Phys. Rev. Lett. 56 (1986) 2453


\bibitem{sklyGodin}   E. K.
 Sklyanin, {\it Separation of variables in the Gaudin model}, J. Sov. Math. 47  (1989)
2473-2488 




 
\bibitem{kunDNLS}
  A. Kundu and B. Basu-mallick, {\it Classical and Quantum integrability of a novel derivative
       NLS model related to quantum group structures.},   { J. Math. Phys.}
     { 34}(1993) 1052 

\bibitem{fadSG} 
  E. K.
Sklyanin , L. A.
 Takhtajan and L. D.
 Faddeev, {\it Quantum inverse problem method I}, Theor. Math. Phys.
40 (1979) 688

 
\bibitem{fadLiu}   L. D.
 Faddeev and O.  Tirkkonen, {\it Connections  of the Liouville model  and XXZ
spin chain}, Nucl. Phys. B453
(1995) 647
\bibitem{chari} V. Chari and A. Presley, {\it Introduction to quantum Groups},
(Cambridge), 1994
\bibitem{kunduPRL99}  A. Kundu,
 {\it Algebraic approach  in unifying   quantum integrable models}
  ,  Phys. Rev. Lett., 82   (1999) 3936
\bibitem{Cal}
 F. Calogero,  {\it Solution of the one-dimensional N-body problem with quadratic
and/or inversely quadratic pair potentials}, J. Math. Phys. 12 (1971)
419-436
\bibitem{Mos}
J. Moser, {\it Three integrable Hamiltonian systems connected with isospectral
deformations} , Adv. Math. 16 (1975) 197-220
\bibitem{Sud}
B. Sutherland,  {\it Exact results for a quantum many-body problem in one
dimension}, Phys. Rev. A5 (1972) 1372


\bibitem{kitaev1} A.
 Yu.
 Kitaev, {\it Fault-tolerant quantum computation by
anyons}, Ann. Phys. 303 (2003) 2-31 
\bibitem{kitaev2} A.
 Yu. Kitaev, {\it Anyons in an exactly solved model and beyond}, Ann.
 Ann. Phys. 321 (2006) 2-111 

\bibitem{suris}
 Yu. B. 
Suris,
 {\it Variational Formulation of Commuting Hamiltonian flows:\ \ Multi-time
 Lagrangian  1-forms},
 arXiv: 1212.3314 [math-ph], v2 (2013)
\bibitem{kunduarXiv12} A. Kundu,
 {\it Unraveling hidden hierarchies and dual structures in an integrable field model}, 
\ \
arXiv: 1201.0627 [nlin.SI], 2012

A.Kundu, {\it Novel  hierarchies and  hidden dimensions in  integrable field
models: theory and application}, J. Phys.: Conf. Ser. 482 (2014) 012022 
 

\bibitem{FadTakhHam} L. D. Faddeev and L. A. Takhtajan, {\it Hamiltonian
methods in the theory of solitons} (Springer Science \& Business Media, 2007)

 \end{thebibliography}
\end{document}